\documentstyle[preprint,tighten,floats,prd,aps]{revtex}
\input epsf

\begin{document}
\preprint{hep-ph/9803211 \hspace{111mm} UAHEP969}
\draft

\title{$O(\alpha^n\alpha_s^m)$ Corrections in
          e$^{+}$e$^{-}$ Annihilation and $\tau$ Decay}

\author{Levan R.\ Surguladze\footnote{levan@gluino.ph.ua.edu}}

\address{Department of Physics \& Astronomy, University of Alabama,
                   Tuscaloosa, AL 35487, USA}
\date{September 1996}
\maketitle
\begin{abstract}
The results of evaluation of mixed QED$\times$QCD corrections to
$R(s)$ in e$^{+}$e$^{-}$ annihilation and $R_{\tau}$ in hadronic decay
of the $\tau$ lepton to $O(\alpha^n\alpha_s^m)$, $m+n\leq 3$, are presented.
The net effect on $\alpha_s(M_Z)$ from the Z decay is only about 0.1\%
and in the $\tau$ decay case the net effect is negligible, as expected.
\end{abstract}


Inclusive processes of the hadronic decays of the Z boson and the 
$\tau$ lepton are in the center of experimental \cite{Zexp,tauexp}
and theoretical \cite{Knirev,tautheor} considerations.
These processes are best suitable for, e.g,
high and low energy determinations of the strong coupling $\alpha_s(s)$,
thus providing one of the crucial tests of the Standard Model (SM).
High experimental and theoretical precisions are important to
to spot any deviations from the SM and thus to find a room for a
possible new physics. The precise $\alpha_s(s)$ is also crucial
for checking of fundamentals of GUT \cite{Lang}.
The up to date results for $\alpha_s$ extracted at different
energies, using various observables
are in satisfactory agreement with the SM \cite{Sop}.
However, the problem of unification of couplings at the GUT scale
remains \cite{Lang}. There has been much attention drown to the
so called $R_{b}-R_{c}$ problem \cite{Lang}.
In the above examples, high experimental and theoretical
precisions  are necessary. A significant progress has been made 
in the last decade in theoretical
evaluation of the above quantities, using perturbation
theory methods \cite{MOR,tautheor}.
To resolve the observed discrepancy between the experiment and the
SM predictions, numerous calculations of contributions beyond the
SM has been done recently (see, e.g., \cite{BSM}).

In the present work we evaluate corrections due to multiple
photon exchange between quarks and between leptons taking place
along the gluon exchange between ``final state'' quarks.
The calculation
has been done to four-loop level corresponding to 
$O(\alpha^n\alpha_s^m)$ ($m+n\leq 3$) corrections.
The objective of this work is to check explicitely that
there are no anomalously large perturbative coefficients
in front of a small $\alpha^n\alpha_s^m$ expansion parameter,
that could make the correction significant.
Such a possibility is not excluded (see, e.g., \cite{mqq}).
Indeed, individual
diagram (or even gauge invariant sets of diagrams) contributions
are often large and only thanks to some delicate cancellations,
they add up to small numbers.

We found that all calculated new corrections are indeed
small and, as expected, they can be neglected at the relevant
level of precision of $O(1\%)$.

The method of calculation is similar to the one used in earlier works
\cite{Rs} and described in ref.\ \cite{RMP}. Here we discuss issues
specific mainly for the present calculation. Known
electroweak and QCD corrections to the Z decay rates are 
summarised in refs.\ \cite{Knirev} and \cite{MOR}, correspondingly.

The decay rate of the Z boson is usually calculated as the imaginary
part of the correlation function of two neutral weak currents of quarks
coupled to Z boson. The decay rate has vector and axial-vector parts,
which generally are different. In the limit of massless
quarks - the approximation used in this work, the vector and axial
parts are identical up to the known factors. 
Therefore, here we calculate only a nonsinglet vector part, or the  
quantity $R(s)$. The total hadronic decay rate $\Gamma_Z$ can then be
obtained straightforwardly, by multiplying the result on the sum
of the squares of vector and axial couplings and by adding
other known corrections from refs.\ \cite{Knirev,MOR}.

The four-loop R(s), including the mixed QCD$\times$QED corrections
of $O(\alpha^n\alpha_s^m)$ ($m+n\leq 3$) has the following
form
\newpage

\begin{eqnarray}
\lefteqn{\hspace{-27mm} 
R(s)=R_0+\frac{\alpha(s)}{\pi}R_1^{(\alpha)}
        +\biggl(\frac{\alpha(s)}{\pi}\biggr)^2R_2^{(\alpha^2)}
        +\biggl(\frac{\alpha(s)}{\pi}\biggr)^3R_3^{(\alpha^3)} }
                                                \nonumber\\
 && \quad
  +\frac{\alpha_s(s)}{\pi}
         \biggl(R_1^{(\alpha_s)}
               +\frac{\alpha(s)}{\pi}R_2^{(\alpha \alpha_s)}
                   +\biggl(\frac{\alpha(s)}{\pi}\biggr)^2
                                R_3^{(\alpha^2 \alpha_s)}\biggr)
                                             \nonumber\\
 && \quad  \hspace{33mm}  
+\biggl(\frac{\alpha_s(s)}{\pi}\biggr)^2
       \biggl(R_2^{(\alpha_s^2)}
            +\frac{\alpha(s)}{\pi}R_3^{(\alpha \alpha_s^2)}
                                                          \biggr)
                                                   \nonumber\\
 && \quad  \hspace{69mm}
+\biggl(\frac{\alpha_s(s)}{\pi}\biggr)^3R_3^{(\alpha_s^3)}.
\label{eq:Rexpans}
\end{eqnarray}
The powers of electromagnetic coupling $\alpha$ and the
strong coupling $\alpha_s$ in the superscripts of $R_i$
indicate correspondingly the number of photon and the number
of gluon exchanges involved in the corresponding multiloop Feynman
graphs. The couplings are renormalized according to the
$\overline{\mbox{\small MS}}$ prescription \cite{MSB}
\begin{equation}
\alpha_s^B=\mu^{2\varepsilon}\alpha_s\biggl[
    1-\frac{\alpha_s}{\pi}
  \biggl(\frac{\beta_0^{(\alpha_s)}}{\varepsilon}
           +\frac{\alpha}{\pi}
         \frac{\beta_1^{(\alpha_s \alpha)}}{2\varepsilon}\biggr)
+\left(\frac{\alpha_s}{\pi}\right)^2 \left(\frac{(\beta_0^{(\alpha_s)})^2}{\varepsilon^2}
-\frac{\beta_1^{(\alpha_s^2)}}{2\varepsilon}\right)\biggr],
\label{eq:AsR}
\end{equation}
\begin{equation}
\alpha^B=\mu^{2\varepsilon}\alpha\biggl[
    1-\frac{\alpha}{\pi}
  \biggl(\frac{\beta_0^{(\alpha)}}{\varepsilon}
           +\frac{\alpha_s}{\pi}
     \frac{\beta_1^{(\alpha \alpha_s)}}{2\varepsilon}\biggr)
+\left(\frac{\alpha}{\pi}\right)^2
   \left(\frac{(\beta_0^{(\alpha)})^2}{\varepsilon^2}
       -\frac{\beta_1^{(\alpha^2)}}{2\varepsilon}\right)\biggr],
\label{eq:AR}
\end{equation}
where the couplings run according to the renormalization group
equations
\begin{equation}
\alpha_s\beta^{\mbox{\scriptsize QCD}}(\alpha,\alpha_s)
    =\mu^2\frac{d\alpha_s}{d\mu^2},
 \hspace{7mm}
\alpha\beta^{\mbox{\scriptsize QED}}(\alpha,\alpha_s)
    =\mu^2\frac{d\alpha}{d\mu^2}
\label{eq:RG}
\end{equation}
and the two-loop renormalization group $\beta$ functions
now include $O(\alpha \alpha_s)$ perturbative terms
\begin{equation}
\beta^{\mbox{\scriptsize QCD}}(\alpha,\alpha_s)
     =-\frac{\alpha_s}{\pi}\biggl(\beta_{0}^{(\alpha_s)}
        +\frac{\alpha}{\pi}\beta_1^{(\alpha_s \alpha)}\biggr)
             -(\frac{\alpha_s}{\pi})^2\beta_1^{(\alpha_s^2)} 
                                         +\cdots,
\label{eq:betaQCD} 
\end{equation}
\begin{equation}
\beta^{\mbox{\scriptsize QED}}(\alpha,\alpha_s)
     =-\frac{\alpha}{\pi}\biggl(\beta_{0}^{(\alpha)}
        +\frac{\alpha_s}{\pi}\beta_1^{(\alpha \alpha_s)}\biggr)
             -(\frac{\alpha}{\pi})^2\beta_1^{(\alpha^2)} 
                                         +\cdots.
\label{eq:betaQED} 
\end{equation}
The known values for one- and two-loop $\beta$ function coefficients
can be found, e.g., in ref.\ \cite{RMP}
\begin{displaymath} 
\beta_0^{(\alpha_s)}=\frac{1}{4}\biggl(\frac{11}{3}C_A
                                       -\frac{4}{3}TN_f\biggr),
\hspace{7mm}
\beta_0^{(\alpha)}= -\frac{1}{3}(N_l+N_F\sum_{f=u,d,...}Q_f^2),
\end{displaymath}

\begin{displaymath}
\beta_1^{(\alpha_s^2)}
=\frac{1}{16}\biggl(\frac{34}{3}C_A^2-\frac{20}{3}C_ATN_f
-4C_FTN_f\biggr), 
\hspace{7mm}
\beta_1^{(\alpha^2)}=-\frac{1}{4}(N_l+N_F\sum_{f=u,d,...}Q_f^4). 
\end{displaymath}
The $O(\alpha \alpha_s)$ coefficients can be found from the
analysis of two-loop graphs and the known two-loop results
\begin{equation}
\beta_1^{(\alpha_s \alpha)}
=-\frac{1}{4}T\sum_{f=u,d,...}Q_f^2, 
\hspace{7mm}
\beta_1^{(\alpha \alpha_s)}=-\frac{1}{4}C_F N_F \sum_{f=u,d,...}Q_f^2. 
\end{equation}
The eigenvalues of the Casimir operators for the adjoint ($N_A=8$)
and the fundamental ($N_F=3$) representations of
SU$_{\mbox{\scriptsize c}}$(3) gauge group are
$C_{A}=3$ and $C_{F}=4/3$. The Dynkin index $T$ for the
fundamental representation is usually chosen $T=1/2$.
$N_l$ is the number of leptons appearing in closed fermion loops
not attached to the gluon propagators. $Q_f$ denotes electric
charge of the quark of flavor $f$ in the units of the electron
charge.

In order to calculate the unknown perturbative terms in Eq.\
(\ref{eq:Rexpans}), we have rederived Eq.\ (6.5) of ref.\
\cite{RMP} - the four-loop expression for R(s) in terms of
perturbative coefficients of the correlation function, its
renormalization constant and the $\beta$ function,
taking into account $O(\alpha^m\alpha_s^n)$ terms.
We have also found the renormalization group constraints,
similar to the ones in Eqs.\ (2.29), (2.30) and (2.32)
of ref.\ \cite{RMP}. These relations are very helpful in testing
the intermediate results.
Contributions from two-, three- and four-loop graphs
were found using the graph-by-graph results of the work
\cite{Rs}. This requred a careful counting of symmetric and
statistical factors for each of the contributing 98 four-loop,
14 three-loop and 2 two-loop Feynman graphs and calculation
of the gauge group weights.

We obtain the following analytical results for the QED and
QED$\times$QCD terms in Eq.\ (\ref{eq:Rexpans}) for the 
standard U(1) and SU$_{\mbox{\scriptsize c}}$(3) gauge groups
within the $\overline{\mbox{\small MS}}$ framework
\begin{equation}
R_1^{(\alpha)}=\frac{9}{4}\sum_{f}Q_f^4,
\label{R1a}
\end{equation}
\begin{equation}
R_2^{(\alpha^2)}=-\frac{9}{32}\sum_{f}Q_f^6
          -\biggl(\frac{33}{8}-3\zeta(3)\biggr)
                              (N_l+3\sum_{j}Q_j^2)\sum_{f}Q_f^4,
\label{R2aa}
\end{equation}
\begin{eqnarray}
\lefteqn{\hspace{-9mm}
R_3^{(\alpha^3)}=-\frac{207}{128}\sum_{f}Q_f^8
      +\biggl(\frac{27}{8}+\frac{39}{4}\zeta(3)-15\zeta(5)\biggr)
                              (N_l+3\sum_{j}Q_j^2)\sum_{f}Q_f^6}
                                                     \nonumber\\
 && \quad \hspace{27mm}
       +\biggl(\frac{151}{18}-\frac{1}{2}\zeta(2)
                          -\frac{19}{3}\zeta(3)\biggr)
                              (N_l+3\sum_{j}Q_j^2)^2\sum_{f}Q_f^4
                                                     \nonumber\\
 && \quad \hspace{49mm}
            -\biggl(\frac{303}{64}-\frac{9}{2}\zeta(3)\biggr)
                              (N_l+3\sum_{j}Q_j^4)\sum_{f}Q_f^4
                                                   \nonumber\\
 && \quad \hspace{59mm}
            +\biggl(\frac{11}{4}-6\zeta(3)\biggr)
                              (N_l+3\sum_{i}Q_i^4)\sum_{f}Q_f^4,
\label{R3aaa}
\end{eqnarray}

\begin{equation}
R_2^{(\alpha \alpha_s)} = - \frac{3}{4}\sum_{f}Q_f^4,
\label{R2as}
\end{equation}

\begin{eqnarray}
\lefteqn{\hspace{-19mm}
R_3^{(\alpha^2 \alpha_s)} = - \frac{207}{32}\sum_{f}Q_f^6
                            -\biggl(\frac{303}{16}-18\zeta(3)\biggr)
                                   \sum_{f}Q_j^2\sum_{f}Q_f^4 }
                                                         \nonumber\\
 && \quad \hspace{19mm}
          +\biggl(\frac{9}{2}+13\zeta(3)-20\zeta(5)\biggr)
                         (N_l+3\sum_{j}Q_j^2)\sum_{f}Q_f^4,
\label{R3aas}           
\end{eqnarray}
\begin{eqnarray}
\lefteqn{
R_3^{(\alpha \alpha_s^2)} = -\biggl[
        \biggl( \frac{519}{16}
                  +\frac{429}{4}\zeta(3)
                                -165\zeta(5) \biggr)
                  -N_f\biggl( \frac{9}{4}
                            +\frac{13}{2}\zeta(3)
                                         -10\zeta(5) \biggr)   
                                                        \biggr]
                                                          \sum_{f}Q_f^4}
                                                   \nonumber\\
 && \quad \hspace{87mm}
   -\biggl(\frac{101}{32}-3\zeta(3) \biggr) \sum_{j}Q_j^2\sum_{f}Q_f^2
                                                       \nonumber\\
 && \quad \hspace{97mm}
   +\biggl( \frac{11}{2}-12\zeta(3) \biggr) \sum_{i}Q_i^2\sum_{j}Q_j^2
\label{R3ass}           
\end{eqnarray}
where summations run over participating quark flavors, normally
- u, d, s, c, b. We keep summation indices different in order
to identify its source topology. For instance, the summation over
$f$ is due to the outer quark loops - the ``final state'' quarks.
The summations over $i$ and $j$ correspond to the inner quark loops
where virtual quarks and leptons can appear. Note that, of course,
leptons can appear only in the case when there is no gluon line
attached to the fermion loop. The last terms in Eqs. (\ref{R3aaa})
and (\ref{R3ass}) are due to the specific four-loop topology
where current operators are inserted in the separate fermion
loops. The two-, three- and four-loop QCD terms are known
and can be found in the original work \cite{Rs}. 
Corrections due to non-vanishing quark masses to these terms
are also known and can be found, for instance, in the review
paper \cite{MOR}. For simplicity, in the present work
we neglect all mass corrections which can trivially be added
to our results.

For five massless quarks and infinitely heavy top quark
we obtain the following numerical result
\begin{eqnarray}
\lefteqn{\hspace{-3mm} 
R(s)=3\sum_{u,d,s,c,b}Q_f^2\biggl\{
       1+0.2652\frac{\alpha(s)}{\pi}
           -3.2747\biggl(\frac{\alpha(s)}{\pi}\biggr)^2
                -2.3539\biggl(\frac{\alpha(s)}{\pi}\biggr)^3}
                                                     \nonumber\\
 && \quad \hspace{49mm}
  +\frac{\alpha_s(s)}{\pi} 
      \biggl[1-0.0884\frac{\alpha(s)}{\pi}
          -0.4089\biggl(\frac{\alpha(s)}{\pi}\biggr)^2\biggr]
                                                     \nonumber\\
 && \quad \hspace{77mm}
  +\biggl(\frac{\alpha_s(s)}{\pi}\biggr)^2
      \biggl[1.409-2.4857\frac{\alpha(s)}{\pi}\biggr]
                                                     \nonumber\\
 && \quad \hspace{109mm}
    -12.805\biggl(\frac{\alpha_s(s)}{\pi}\biggr)^3
                                              \biggr\}.
\label{eq:Rnum}
\end{eqnarray}
One can see that the calculated corrections are very small.
The largest corrections beyond the leading QED term -
$\sim \alpha^2$,
$\sim \alpha_s \alpha$ and $\sim \alpha_s^2 \alpha$
have a same sign and they add up to the total effect
of about 0.1 MeV on the Z width and only 0.1\% on $\alpha_s$.
The other calculated corrections are clearly negligible
for any phenomenological applications.
The calculated terms of $O(\alpha^2)$ and $O(\alpha \alpha_s)$
confirm known results \cite{Kat}.

Let us use the calculated corrections on the Z width and 
obtain the similar terms for the quantity
\begin{equation}
R_{\tau}=\frac{\Gamma(\tau^{-}\rightarrow\nu_{\tau}+\mbox{hadrons})}
       {\Gamma(\tau^{-}\rightarrow\nu_{\tau}e^{-}\overline{\nu}_{e})}
\label{eq:Rtaudef}
\end{equation}
in the $\tau$ lepton hadronic decay process.
Following the well known procedure \cite{Braat}, we calculate
the perturbative part as an integral (see, e.g., \cite{RMP} for details)
\begin{equation}
R_{\tau}^{\mbox{\scriptsize{pert}}}=\frac{3i}{8\pi}\int_{C}\frac{ds}
{M_{\tau}^2}
\biggl(1-\frac{s}{M_{\tau}^2}\biggr)^2
\biggl(1+2\frac{s}{M_{\tau}^2}\biggr)\Pi(s),
\label{eq:Contourint}
\end{equation}
where $\Pi(s)$ is the transverse part of the correlation function
of weak currents of quarks coupled to W boson.
The countour $C$ is the circle of radius $|s|=M_{\tau}^2$.
In the integrand we substitute the $\alpha_s(s)$ and 
$\alpha(s)$ by their expansion in terms of 
$\alpha_s(M_{\tau})$ and $\alpha(M_{\tau})$ using the 
equations
\begin{eqnarray}
\lefteqn{\hspace{-9mm}\frac{\alpha_s(s)}{\pi}
                         =\frac{\alpha_s(M_{\tau})}{\pi}
      +\biggl(\frac{\alpha_s(M_{\tau})}{\pi}\biggr)^2
  \biggl[
     \beta_0^{(\alpha_s)}\log\frac{M_{\tau}^2}{s}
   + \biggl(\frac{\alpha(M_{\tau})}{\pi}\biggr)
       \beta_1^{(\alpha_s\alpha)}\log\frac{M_{\tau}^2}{s}
                                     \biggr] }
                                                     \nonumber\\
 &&  \quad \hspace{43mm}
      +\biggl(\frac{\alpha_s(M_{\tau})}{\pi}\biggr)^3
      \biggl[\beta_1^{(\alpha_s^2)}\log\frac{M_{\tau}^2}{s} 
      +(\beta_0^{(\alpha_s)})^2\log ^2\frac{M_{\tau}^2}{s}\biggr],
\label{eq:Astransform}
\end{eqnarray}
\begin{eqnarray}
\lefteqn{\hspace{-9mm}\frac{\alpha(s)}{\pi}
                         =\frac{\alpha(M_{\tau})}{\pi}
      +\biggl(\frac{\alpha(M_{\tau})}{\pi}\biggr)^2
  \biggl[
     \beta_0^{(\alpha)}\log\frac{M_{\tau}^2}{s}
   + \biggl(\frac{\alpha_s(M_{\tau})}{\pi}\biggr)
       \beta_1^{(\alpha\alpha_s)}\log\frac{M_{\tau}^2}{s}
                                     \biggr] }
                                                     \nonumber\\
 &&  \quad \hspace{43mm}
      +\biggl(\frac{\alpha(M_{\tau})}{\pi}\biggr)^3
      \biggl[\beta_1^{(\alpha^2)}\log\frac{M_{\tau}^2}{s} 
      +(\beta_0^{(\alpha)})^2\log ^2\frac{M_{\tau}^2}{s}\biggr].
\label{eq:Atransform}
\end{eqnarray}
Then we evaluate contour integral and express $R_{\tau}$
in terms of the perturbative coefficients $R_i$ in Eq.\
(\ref{eq:Rexpans}), making obvious substitution
$\sum_{f}Q_f^2 \rightarrow |V_{ud}|^2+|V_{us}|^2$ and
omitting terms coming from the four-loop topologies, where
current insertions are placed in the separate fermion loops.
Thus for the perturbative part of the $R_{\tau}$ we obtain
the following $\overline{\mbox{\small MS}}$ analytical result
\begin{eqnarray}
\lefteqn{ 
R_{\tau}^{\mbox{\scriptsize{pert}}}=
   3(\mid V_{ud}\mid^2+\mid V_{us}\mid^2)
   \biggl\{
    1+\frac{\alpha(M_{\tau})}{\pi}r_1^{(\alpha)}
        +\biggl(\frac{\alpha(M_{\tau})}{\pi}\biggr)^2 r_2^{(\alpha^2)}
        +\biggl(\frac{\alpha(M_{\tau})}{\pi}\biggr)^3 r_3^{(\alpha^3)} }
                                                \nonumber\\
 && \quad \hspace{53mm}
  +\frac{\alpha_s(M_{\tau})}{\pi}
         \biggl(r_1^{(\alpha_s)}
               +\frac{\alpha(M_{\tau})}{\pi}r_2^{(\alpha \alpha_s)}
                   +\biggl(\frac{\alpha(M_{\tau})}{\pi}\biggr)^2
                                r_3^{(\alpha^2 \alpha_s)}\biggr)
                                             \nonumber\\
 && \quad  \hspace{83mm}  
+\biggl(\frac{\alpha_s(M_{\tau})}{\pi}\biggr)^2
       \biggl(r_2^{(\alpha_s^2)}
            +\frac{\alpha(M_{\tau})}{\pi}r_3^{(\alpha \alpha_s^2)}
                                                          \biggr)
                                                   \nonumber\\
 && \quad  \hspace{114mm}
+\biggl(\frac{\alpha_s(M_{\tau})}{\pi}\biggr)^3 r_3^{(\alpha_s^3)}
                                                             \biggr\},
\label{eq:Rtauexp}
\end{eqnarray}
where for the 
standard U(1) and SU$_{\mbox{\scriptsize c}}$(3) gauge groups
within the $\overline{\mbox{\small MS}}$ framework we obtain
\begin{equation}
r_1^{(\alpha)}=\frac{3}{4}\sum_{f}Q_f^2,
\label{r1a}
\end{equation}
\begin{equation}
r_2^{(\alpha^2)}=-\frac{3}{32}\sum_{f}Q_f^4
          -\biggl(\frac{85}{48}-\zeta(3)\biggr)
                              (N_l+3\sum_{j}Q_j^2)\sum_{f}Q_f^2,
\label{r2aa}
\end{equation}
\begin{eqnarray}
\lefteqn{\hspace{-9mm}
r_3^{(\alpha^3)}=-\frac{69}{128}\sum_{f}Q_f^6
      +\biggl(\frac{235}{192}+\frac{13}{4}\zeta(3)-5\zeta(5)\biggr)
                              (N_l+3\sum_{j}Q_j^2)\sum_{f}Q_f^4}
                                                     \nonumber\\
 && \quad \hspace{27mm}
       +\biggl(\frac{3935}{864}-\frac{1}{3}\zeta(2)
                          -\frac{19}{6}\zeta(3)\biggr)
                              (N_l+3\sum_{j}Q_j^2)^2\sum_{f}Q_f^2
                                                     \nonumber\\
 && \quad \hspace{59mm}
            -\biggl(\frac{15}{8}-\frac{3}{2}\zeta(3)\biggr)
                              (N_l+3\sum_{j}Q_j^4)\sum_{f}Q_f^2,
\label{r3aaa}
\end{eqnarray}

\begin{equation}
r_2^{(\alpha \alpha_s)} = - \frac{1}{4}\sum_{f}Q_f^2,
\label{r2as}
\end{equation}

\begin{eqnarray}
\lefteqn{\hspace{-7mm}
r_3^{(\alpha^2 \alpha_s)} = - \frac{69}{32}\sum_{f}Q_f^4
+\biggl(\frac{235}{144}+\frac{13}{3}\zeta(3)-\frac{20}{3}\zeta(5)\biggr)
                         (N_l+3\sum_{j}Q_j^2)\sum_{f}Q_f^2 }
                                                         \nonumber\\
 && \quad \hspace{69mm}
          -\biggl(\frac{15}{2}-6\zeta(3)\biggr)
                         \sum_{j}Q_j^2\sum_{f}Q_f^2,
\label{r3aas}           
\end{eqnarray}
\begin{equation}
r_3^{(\alpha \alpha_s^2)} = 
        -\biggl( \frac{605}{64}
                  +\frac{117}{4}\zeta(3)
                                -45\zeta(5) \biggr)
   -\biggl(\frac{5}{4}-\zeta(3) \biggr) \sum_{j}Q_j^2.
\label{r3ass}           
\end{equation}
The summation index $f$ runs over $u, d, s$ ``final state''
quark flavors appearing in the outer fermionic loops and the
summation over $j$ corresponds inner (virtual) fermionic
loops. We take $j=u,d,s$. For the number of virtual leptons
in the inner fermionic loops, we will take $N_l=2$ ($e,\mu$).
In fact, the $\tau$-lepton can also appear in the virtual loops.
However, within our approximation, masses of all quarks and leptons,
including virtual ones are neglected. On the other hand, the $\tau$ mass
cannot be neglected at this scale. Therefore, the effect
of the virtual $\tau$ lepton is left out of our consideration.
This effect can be evaluated via exact evaluation of Feynman graphs
with massive loops. The corresponding numerical effect is expected to be
small.
The two-, three- and four-loop QCD terms are known
and can be found in \cite{Rtau}.
We obtain the following numerical expression
\begin{eqnarray}
\lefteqn{\hspace{-3mm}
R_{\tau}^{\mbox{\scriptsize{pert}}}=
   3(\mid V_{ud}\mid^2+\mid V_{us}\mid^2) \biggl\{
       1+0.5\frac{\alpha(s)}{\pi}
           -1.5376\biggl(\frac{\alpha(s)}{\pi}\biggr)^2
                +1.9041\biggl(\frac{\alpha(s)}{\pi}\biggr)^3}
                                                     \nonumber\\
 && \quad \hspace{49mm}
  +\frac{\alpha_s(s)}{\pi} 
      \biggl[1-0.16667\frac{\alpha(s)}{\pi}
          -0.7990\biggl(\frac{\alpha(s)}{\pi}\biggr)^2\biggr]
                                                     \nonumber\\
 && \quad \hspace{77mm}
  +\biggl(\frac{\alpha_s(s)}{\pi}\biggr)^2
      \biggl[5.202+1.2778\frac{\alpha(s)}{\pi}\biggr]
                                                     \nonumber\\
 && \quad \hspace{109mm}
    +26.366\biggl(\frac{\alpha_s(s)}{\pi}\biggr)^3
                                              \biggr\}.
\label{eq:Rtaunum}
\end{eqnarray}
As one can see, the higher order QED and QED$\times$QCD corrections
are very small. Even the largest ones of $O(\alpha \alpha_s)$
and $O(\alpha \alpha_s^2)$ have similar size but opposite signs
and they cancel each other along the $O(\alpha^2)$ correction.
Thus, what remains is the leading $O(\alpha)$ correction that has
some numerical relevance. All other corrections are clearly negligible.

Summarising, we note that the calculated higher order
QED and mixed  QED$\times$QCD corrections to the hadronic decay
rates of the Z boson and the $\tau$ lepton can be safely neglected
in present phenomenological applications.

\acknowledgements

The author is grateful to L.\ Clavelli and P.\ Coulter for discussions.
This work was supported by the U.S. Department of Energy under
grant No.\ DE-FG02-96ER-40967.


\begin{references}


\bibitem{Zexp} D.\ Schaile, in {\it Proceedings of the 27th
               International Conference on High Energy Physics}
               (Glasgow, Scotland, 20-27 July, 1994), p.\ 27.  
\bibitem{tauexp} L.\ Duflot,
                 Nucl.\ Phys.\ Proc.\ Suppl.\ {\bf 40}, 37 (1995).
\bibitem{Knirev} B.\ A.\ Kniehl,
                 Int.\ J.\ Mod.\ Phys.\ {\bf A 10}, (1995) 443;
                 Phys.\ Rep.\  {\bf 240}, 211 (1994).
\bibitem{tautheor} W.\ J.\ Marciano, in {\it Proceedings of the
                   Third Workshop
                   on Tau Lepton Physics} 
                   (Montreaux, Switzerland, September 19-22, 1994);
                   Preprint BNL-61141, 1994;
                   A.\ Pich,
                   Nucl.\ Phys.\ Proc.\ Suppl.\ {\bf 39 BC}, 326 (1995).
\bibitem{Lang} P.\ Langacker, in {\it Precision Tests of the Standard
               Electroweak Model}, ed. by P.\ Langacker
               (World Scientific, 1994) 
\bibitem{Sop} D.\ E.\ Soper, 
              in {\it Proceedings of the XXXth Rencontre de Moriond,
              QCD and High Energy Hadronic Interactions,
              Les Arcs, Savoie, France,} 18-25 March 1995,
              edited by J.\ Tran Thanh Van
              (Editions Frontieres, France, 1995),
              p.\ 615;
              S.\ Bethke, {\it ibid.} p.\ 213. 
\bibitem{MOR} D.\ E.\ Soper and L.\ R.\ Surguladze,
              in {\it Proceedings of the XXXth Rencontre de Moriond,
              QCD and High Energy Hadronic Interactions,
              Les Arcs, Savoie, France,} 18-25 March 1995,
              edited by J.\ Tran Thanh Van
              (Editions Frontieres, France, 1995),
              p.\ 71.
\bibitem{BSM} L.\ Clavelli, Mod.\ Phys.\ Lett.\  {\bf A 10}, 949 (1995);
              J.\ D.\ Wells, C.\ Kolda and G.\ L.\ Kane,
              Phys.\ Lett.\ {\bf B 338}, 219 (1994);
              J.\ T.\ Liu and D.\ Ng, {\it ibid.} {\bf 342}, 262 (1995);
              L.\ R.\ Surguladze and L.\ J.\ Clavelli,
              Phys.\ Rev.\ Lett.\ {\bf 78}, 1632 (1997).
\bibitem{mqq} L.\ R.\ Surguladze and F.\ V.\ Tkachov, 
           Nucl.\ Phys.\ {\bf B 331}, 35 (1990).
\bibitem{Rs} L.\ R.\ Surguladze and M.\ A.\ Samuel,
                  Phys.\ Rev.\ Lett.\ {\bf 66}, 560 (1991);
                  see also 
                  Phys.\ Lett.\ {\bf B 309}, 157 (1993).
\bibitem{RMP} L.\ R.\ Surguladze and M.\ A.\ Samuel,
                   Rev.\ Mod.\ Phys.\ {\bf 68}, 256 (1996).
\bibitem{MSB} G.\ t'~Hooft, Nucl.\ Phys.\ {\bf B 61}, 455 (1973);
              W.~Bardeen, A.~Buras, D.~Duke and T.~Muta,
              Phys.\ Rev.\ {\bf D 18}, 3998 (1978).
\bibitem{Kat} A.\ L.\ Kataev, Phys.\ Lett.\ {\bf B 287}, 209 (1992).
\bibitem{Braat} E.\ Braaten, Phys.\ Rev.\ Lett.\ {\bf 60}, 1606 (1988).
\bibitem{Rtau} M.\ A.\ Samuel and L.\ R.\ Surguladze,
               Phys.\ Rev.\ {\bf D 44}, 1602 (1991). 
\end{references}
\end{document}